\documentclass[aps,prl,twocolumn,tightenlines,superscriptaddress]{revtex4-1}
\usepackage{epsfig,rotating}
\usepackage{multirow}
\usepackage{amsbsy}
\usepackage{stmaryrd} 

\newcommand{\surf}{{\textrm{{\tiny $\ell$}}}}
\newcommand{\surfoil}{{\textrm{{\tiny $\ell$-oil}}}}

\newcommand{\bulk}{{\textrm{\tiny b}}}
\newcommand{\bulkoil}{{\textrm{\tiny b-oil}}}

\newcommand{\DD}{{\textrm{\tiny d}}}
\newcommand{\YM}{{\textrm{\tiny 2M}}}
\newcommand{\oneoil}{{\textrm{\tiny 1,oil}}}
\newcommand{\onewater}{{\textrm{\tiny 1,wat}}}
\newcommand{\twooil}{{\textrm{\tiny 2,oil}}}
\newcommand{\twowater}{{\textrm{\tiny 2,wat}}}
\newcommand{\onebulkoil}{{\textrm{\tiny 1,b-oil}}}
\newcommand{\onesurfoil}{{\textrm{\tiny 1,$\ell$-oil}}}
\newcommand{\onebulkwater}{{\textrm{\tiny 1,b-wat}}}
\newcommand{\onesurfwater}{{\textrm{\tiny 1,$\ell$-wat}}}
\newcommand{\oneMn}{{\textrm{\tiny 1,aq}}}

\begin{document}
\title{Explicit calculation of nuclear magnetic resonance relaxation rates in small pores to elucidate molecular scale fluid dynamics}

\author{D.\ A.\ Faux}
     \affiliation{Department of Physics, University of Surrey,
       Guildford, Surrey GU2 7XH, United Kingdom}
\author{P.\ J.\ McDonald}
\affiliation{Department of Physics, University of Surrey,
	Guildford, Surrey GU2 7XH, United Kingdom}

   \date{\today}

\begin{abstract}
A model linking the molecular-scale dynamics of fluids confined to nano-pores to nuclear magnetic resonance (NMR) relaxation rates is proposed. The model is fit to experimental NMR dispersions for water and oil in an oil shale assuming that each fluid is characterised by three time constants and L\'{e}vy statistics.  Results yield meaningful and consistent intra-pore dynamical time constants, insight into diffusion mechanisms and pore morphology. The model is applicable to a wide range of porous systems and advances NMR dispersion as a powerful tool for measuring nano-porous fluid properties.
\end{abstract}

\maketitle

\begin{figure}[tbh!]			
	\centering
	\includegraphics[width=8.6 cm, height=6.0cm, trim={0cm 0 1cm 1cm},clip]{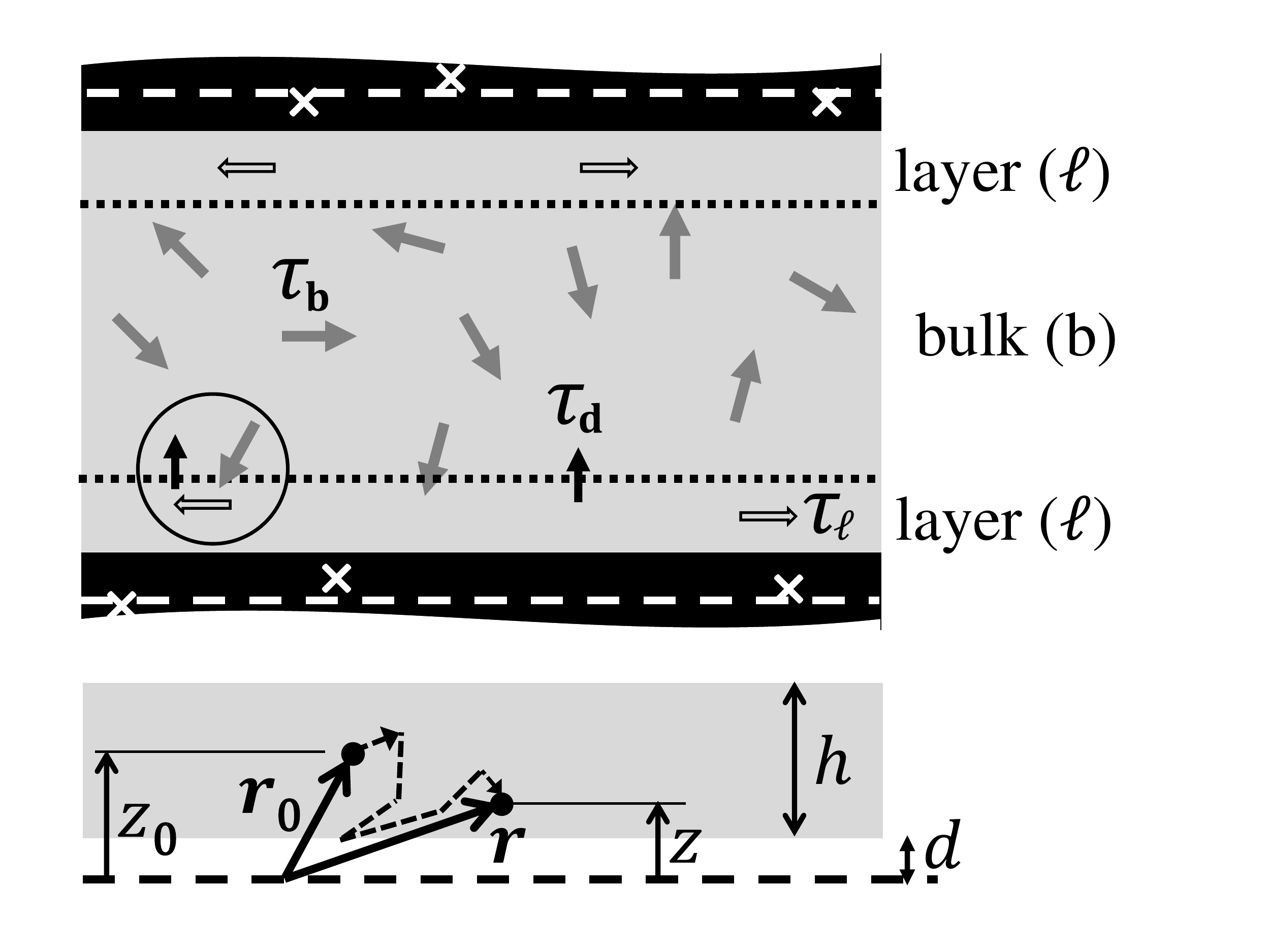}\\[-5 mm]
	\caption{At top, a model quasi-two-dimensional pore shows fluid (gray) confined by walls (black).  Surface diffusion, desorption and bulk diffusion events are characterised by    time constants $\tau_\surf$, $\tau_\DD$ and $\tau_\bulk$ respectively. Rare paramagnetic impurities are indicated by white crosses modelled by a layer of uniform density (white dashed line). The circle indicates a diffusion mechanism consistent with $\tau_\DD\!\approx \!\tau_\surf$ (see text). At bottom, spin pair vectors used in the theory are shown for a mobile spin (black circle). A volume of fluid (gray $\equiv$  surface or bulk fluid) has thickness $h$  located distance $d$ from the paramagnetic layer. \\[-0.7cm] }
	\label{Fig1_model}
\end{figure}

Understanding molecular-scale fluid dynamics in micro- and meso-porous materials is central to understanding a wide range of industrially-important materials and processes: rocks for petroleum engineering; zeolites for catalysis; calcium-silicate-hydrates for concrete construction; bio-polymers for food production to name but a few.  A molecular-scale model of fluid in a pore is depicted in Fig.~\ref{Fig1_model}. In this general picture, one considers fluid within the body  of the pore and a surface layer of fluid at the pore wall. The pore body fluid behaves much as a bulk fluid, free to diffuse in three dimensions with motion characterised by a correlation time $\tau_\bulk$. The surface layer diffuses in just two dimensions (2D) with motion characterised by a slower correlation time $\tau_\surf$. Molecular exchange is envisaged between the surface layer and the bulk fluid characterised by a desorption time $\tau_\DD$ and a corresponding adsorption time linked to $\tau_\DD$ by the requirements of mass balance. This model therefore simplifies the complex intra-pore   dynamics of real fluids to three characteristic time constants,$\tau_\bulk$, $\tau_\surf$ and $\tau_\DD$.  Aspects of this general model, henceforth referred to as the $3\tau$ model, are widely used throughout literature \cite{Zavada.1999,Kimmich.2002,Barberon.2003,McDonald.2005,Korb.2011,Korb.2014, Faux.2013,Faux.2015}.

Nuclear magnetic resonance (NMR) relaxation analysis is a uniquely powerful tool to access molecular correlation times of fluids in porous media \cite{Zavada.1999,Kimmich.2002,Barberon.2003,McDonald.2005,Korb.2011,Korb.2014, Faux.2013,Faux.2015,Fleury.2016}. It is rivalled only by small-angle scattering techniques, especially with neutrons, but has the advantage of being widely available using laboratory-scale equipment. Two NMR relaxation methods are especially valuable. NMR relaxation dispersion (NMRD) measurement of the frequency dependence of the nuclear (usually $^1$H) spin-lattice relaxation time ($T_1$) of fluid molecules in the low-frequency range (kHz to MHz) is sensitive to fluid correlation times. Second, the $T_1$--$T_2$ correlation experiment measures the ratio of $T_1$ to the nuclear spin-spin relaxation time $T_2$. This is especially sensitive to different relaxation mechanisms.

However, for the NMR methods to be useful, a model is required to link fluid molecular dynamics in pores to NMR relaxation rates. Several models have been proposed (for example, \cite{Zavada.1999,Kimmich.2002,Godefroy.2001,Faux.2013,Korb.2014}) but that which builds most successfully on the general dynamics of the model illustrated in Fig.~\ref{Fig1_model} in terms of fitting experimental data is due to Korb and co-workers \cite{Godefroy.2001,Barberon.2003,McDonald.2005,Korb.2011,Korb.2014}.  Korb's model reproduces the fundamental form of the $T_1$ dispersion curve at low frequency in most systems and predicts the $T_1/T_2$ ratio. The model supposes that the dominant relaxation mechanism involves repeated encounters of the diffusing surface layer molecules with static surface relaxation sites, most typically paramagnetic impurities. Korb's model identifies 3 key parameters: two are the correlation times $\tau_\surf$ and $\tau_\DD$ of the general model (Fig.~\ref{Fig1_model}); the third is a frequency-independent bulk-fluid spin-lattice relaxation time $T_{1,\textrm{\tiny b}}$. It is roughly linked to $\tau_\bulk$. From these parameters Korb produces, first, an approximate surface-diffusion-driven temporal nuclear magnetic correlation function $G(t)$ and, second, the relaxation rates. With time and varied application, two limitations of the Korb model have become apparent. The first is that the physical parameters $\tau_\surf$ and $\tau_\DD$ required to fit experimental data are remarkably uniform, typically about 1~ns and 1--10~$\mu$s, respectively. A lack of sensitivity to the diversity of experimental systems studied seems to imply an underlying problem. The second is that it is very hard to justify the correlation times in terms of physics and chemistry. Surface molecules must undergo $10^3$--$10^5$ surface hops across the pore surface {\em without desorbing}. That the molecules must be both ``sticky" and ``non-sticky" at the same time is seemingly contradictory.

In this letter, we propose a model of NMR relaxation of fluids in pores that: (i) {\em preserves} the presumed fluid dynamics captured in the 3$\tau$ model (Fig.~\ref{Fig1_model}); (ii)   achieves {\em improved} fits to experimental data {\em and} (iii) predicts physically-realistic parameters. The model has exactly the same number of adjustable parameters as the Korb model. The model also retains the essential relaxation mechanism of Korb (surface interactions). However, three advances conspire to have critical and profound effect on the outcomes. The three key advances included in our model are as follows.  First, we assume that the paramagnetic relaxation centres are embedded {\em in} the pore wall whereas the Korb model assumes them to reside {\em on} the pore wall.  From this the correlation function $G(t)$ is calculated explicitly and found to vary as $t^{-2}$ in the long-time limit. The Korb model has the functional form $t^{-1}$ in the long-time limit due to  mobile spins and paramagnetic impurities lying in the same 2D plane. In the Korb model, an {\em approximate} construction for $G(t)$ is obtained by combining this long-time dependence with an assumption that molecules desorb from the surface (and do not return).  This ensures that the mathematics is tractable but leads to physically-unrealistic desorption times when data is fit.  By contrast, we admit full L\'{e}vy walk statistics into the model to capture re-adsorption of desorbed molecules. Finally, we integrate $G(t)$ across the full width of the pore in order to properly calculate the frequency dependence of $T_{1,\textrm{\tiny b}}$. This corresponds to recognising that $\tau_\bulk$ is the proper constant of the bulk fluid dynamics, not $T_{1,\textrm{\tiny b}}$.

We have applied our model to varied published experimental data sets of interest to different user communities. We find that best fit parameters vary between experimental systems in a coherent fashion. Here we exemplify the model with analysis of data from an oil shale: a complex two-fluid system of topical interest. The model is most profound here: Korb's model interprets the data as showing that the shale is {\em water} wetting; our model predicts {\em oil} wetting.

A theoretical analysis is now presented which determines  $T_1^{-1}$ and $T_2^{-1}$  based on $3\tau$ dynamics (Fig.~\ref{Fig1_model}).
$P({\bf r},t \cap  {\bf r}_0)$ is the probability density function describing the probability that a spin (either in the surface layer or bulk) is located at  ${\bf r}_0$ relative to an electronic paramagnetic spin at $t\!=\!0$ and ${\bf r}$ at time $t$, as in Fig.~\ref{Fig1_model}. $P({\bf r},t \cap  {\bf r}_0)$ may be written using cylindrical coordinates as
\begin{eqnarray}
P({\bf r},t \:\cap \: {\bf r}_0) & = & N \: P(\pmb{\rho},t \: | \: \pmb{\rho}_0) \: P(z,t \: | \: z_0) \label{Prho}
\end{eqnarray}
where $N$ is the number of paramagnetic spins per unit volume, $P(\pmb{\rho},t \: | \: \pmb{\rho}_0)$ describes the probability that a spin pair has an in-plane displacement $\pmb{\rho}$ at time $t$  {\em given} the displacement was $\pmb{\rho}_0$ at $t\!=\!0$ and is described by L\'{e}vy walk statistics via the transform
\begin{eqnarray}
P(\pmb{\rho},t \: | \: \pmb{\rho}_0) & = & \frac{1}{4 \pi^2} \int e^{-Dtk^\alpha}  e^{ i {\bf k} \cdot \pmb{\rho} } e^{-i {\bf k} \cdot \pmb{\rho_0}}  d^2 {\bf k} \label{eq:fourierintegral}
\end{eqnarray}
where ${\bf k}$ is an in-plane Fourier variable and $0\!<\!\alpha \!\leq\! 2$ is the L\'{e}vy parameter.  If $\alpha\!=\!2$, Eq.~(\ref{eq:fourierintegral}) represents the transform of a Gaussian function and normal Fickian diffusion is recovered.  If $\alpha\!<\!2$, the probability distribution possesses power-law tails providing enhanced probability density in the wings of the distribution. $P(z,t \: | \: z_0)$ is obtained as a solution to the diffusion equation with reflective boundaries as \cite{Bickel.2007}
\begin{eqnarray}
\!\!\!\!\!\! P(z,t \:| \:z_0) & =&  \frac{1}{h} \left[1+2 \: \sum_{p=1}^\infty e^{-D p^2  \pi^2 t/h^2} c_p(z) \: c_p(z_0) \right]  \label{Pzeta}
\end{eqnarray}
where $c_p(z)\!=\!\cos(p\pi(z\!-\!d)/h)$ and the mobile spins are confined to the region $d\!<z\!<d\!+\!h$ as in Fig.~\ref{Fig1_model}.  

The dipolar correlation function $G(t)$ is \cite{Faux.2013,Abragam}
\begin{eqnarray}
G(t) & = & \frac{4 \pi}{5} \int \!\!\! \int
\sum_{M=-2}^{2}  \frac{Y_{\YM}(\rho_0, \phi_0, z_0) \: Y^*_{\YM}(\rho, \phi, z)}{ (\rho^2_0 + z_0^2)^{3/2}  \: (\rho^2 + z^2)^{3/2}}  \nonumber \\[1mm]
&& \times  \: P({\bf r},t \:\cap \: {\bf r}_0) \: \:d^3 {\bf r}_0 \:\: d^3 {\bf r}  \label{eqn:G2}
\end{eqnarray}
where the $Y$ are the spherical harmonic functions of degree 2 where the asterisk represents the complex conjugate.  The powder average has been taken reflecting the (assumed) uniform random orientation of pores in experimental samples \cite{Abragam,Faux.2013}. Substitution of Eqs.~(\ref{Prho})--(\ref{Pzeta}) into Eq.~(\ref{eqn:G2}),  application of the Jacobi-Anger expression followed by volume integrations finally yields
\begin{eqnarray}
\!\!\!\!G(t) & = & \frac{2 N}{5 \delta^3 \Delta}   \int_0^\infty \!\!\!\! e^{-t \kappa^\alpha/6 \tau} \kappa   \Big[ H(\kappa)\! +\! 
2 \sum_{p=1}^\infty e_p (t,\kappa) \Big] d \kappa \label{eqn:G4}
\end{eqnarray}
where $\kappa\!=\!k\delta$ is a dimensionless Fourier variable and
\begin{eqnarray}
H(\kappa) & = & \frac{5 \pi}{3}  \left( e^{\kappa \Delta} - 1 \right)^2  e^{-2\kappa (\Delta+\eta)}  \label{eqn:Hkappa} \\[2mm]
\!\!\!\!e_p(t,\kappa) & = &  \frac{5 \pi \kappa^4 \Delta^4 \; \left[ e^{\kappa \Delta} \!-\! (-1)^p  \right]^2 }{3 \; \left[\kappa^2 \Delta^2 + p^2 \pi^2 \right]^2 \: e^{2\kappa (\Delta+\eta)}} e^{-p^2 \pi^2  t/6 \Delta^2 \tau}.
\label{eqn:Ckappa}
\end{eqnarray}
The dimensionless distances $\eta$ and $\Delta$ are $d/\delta$ and $h/\delta$ respectively where $\delta$ is a convenient molecular-scale distance taken as 0.27~nm, the approximate inter-molecular spin-spin distance in water.  $\delta$ also links $\tau_\surf$ and $\tau_\bulk$ to their diffusion coefficient via $D\!=\!\delta^2/6\tau$.

Finally, the spectral density function $J(\omega)$ is obtained from the Fourier transformation of $G(t)$ allowing $T_1^{-1}$ and $T_2^{-1}$ to be found as follows \cite{Abragam,McDonald.2005}
\begin{eqnarray}
J(\omega) & = & 2 \int^\infty_{0} \!\!  G(t) \: \cos \omega t \: dt \label{Jeqn} \\[2mm]
T_1^{-1} =  &&  \frac{1}{3} \beta \left[  7 J(\omega_\sigma) + 3 J(\omega_p)\right] \label{T1sigma} \\[2mm]
T_2^{-1} =  &&  \frac{1}{6} \beta \left[  4 J(0) + 13 J(\omega_\sigma) + 3 J(\omega_p)\right]. \label{T2sigma}
\end{eqnarray}
Here $ \beta\!=\!\left( \mu_0 / 4\pi \right)^2 \gamma_p^2  \gamma_\sigma^2 \: \hbar^2 S(S+1)$,  $\gamma_\sigma$ ($\gamma_p$) is the gyromagnetic ratio for the paramagnetic impurity (proton) and $S \!= \!\frac{5}{2}$ for Mn$^{2+}$ or Fe$^{3+}$.  $\omega_p$ is the Larmor frequency of a proton in the applied static field and $\omega_\sigma \!= \! 658.21 \omega_p$. 

\begin{figure}[tbh!]
	\begin{center}
		\includegraphics[width=8.6 cm, height=5.5 cm, trim={1cm 1.5cm 1cm 1.0cm},clip]{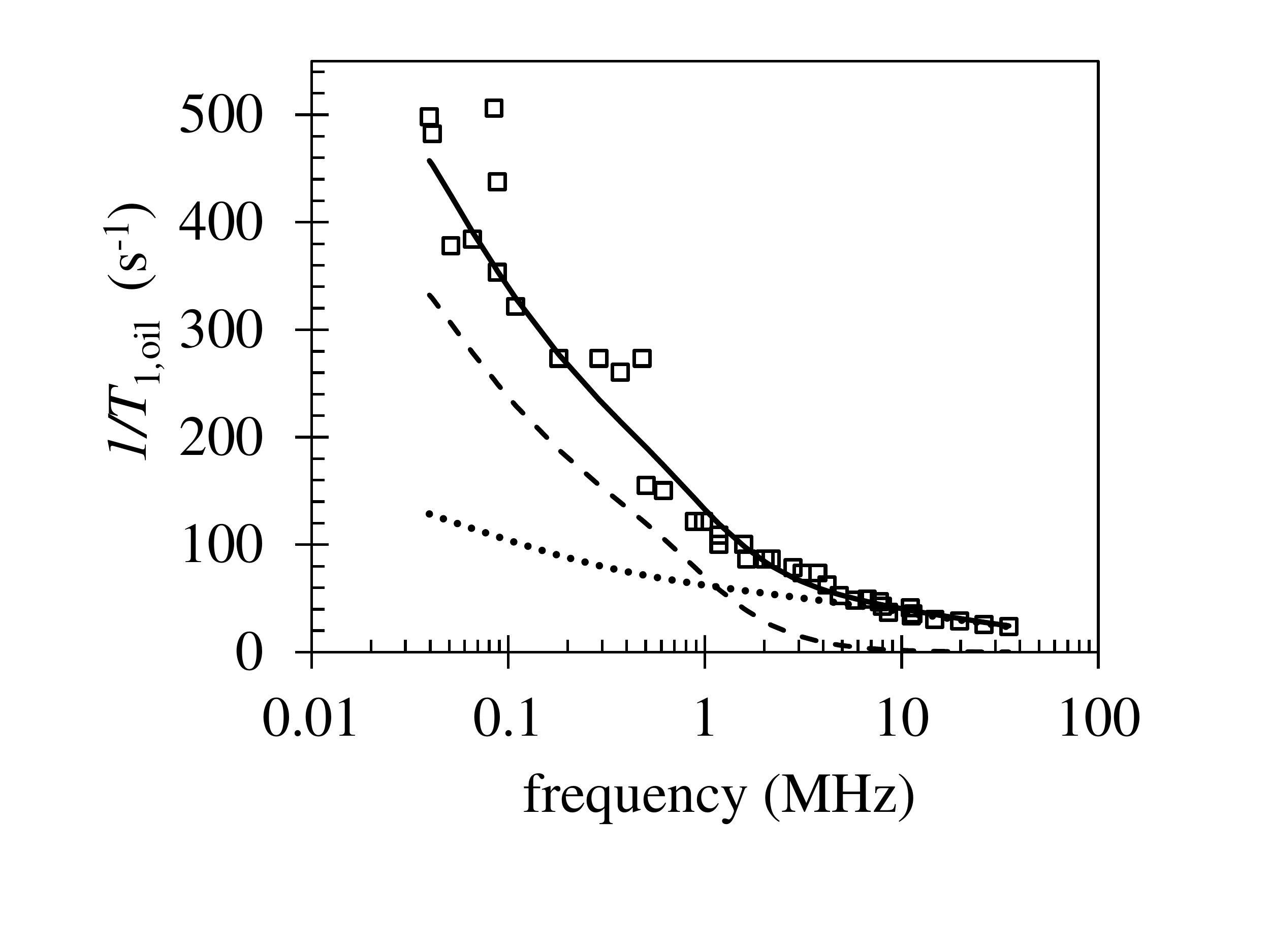} \\[-0.7 cm]
		\caption["Short" caption without tikz code]{$T^{-1}_\oneoil$ is presented as a function of frequency for oil in an oil shale. The experimental data ($\boxempty$) is from Ref.~\cite{Korb.2014}.  The theoretical curve (\raisebox{.8ex}{\line(1,0){20}}) is composed of a bulk ($\cdots$) and surface (\raisebox{.8ex}{\line(1,0){5}~~\line(1,0){5}}) contributions.\\[-0.7cm]  }
		\label{Fig2:shale_oil}
	\end{center}
\end{figure}

\begin{figure}[tbh!]
	\begin{center}
		\includegraphics[width=7.6 cm, height=5.5 cm, trim={1cm 1cm 0cm 1.0cm},clip]{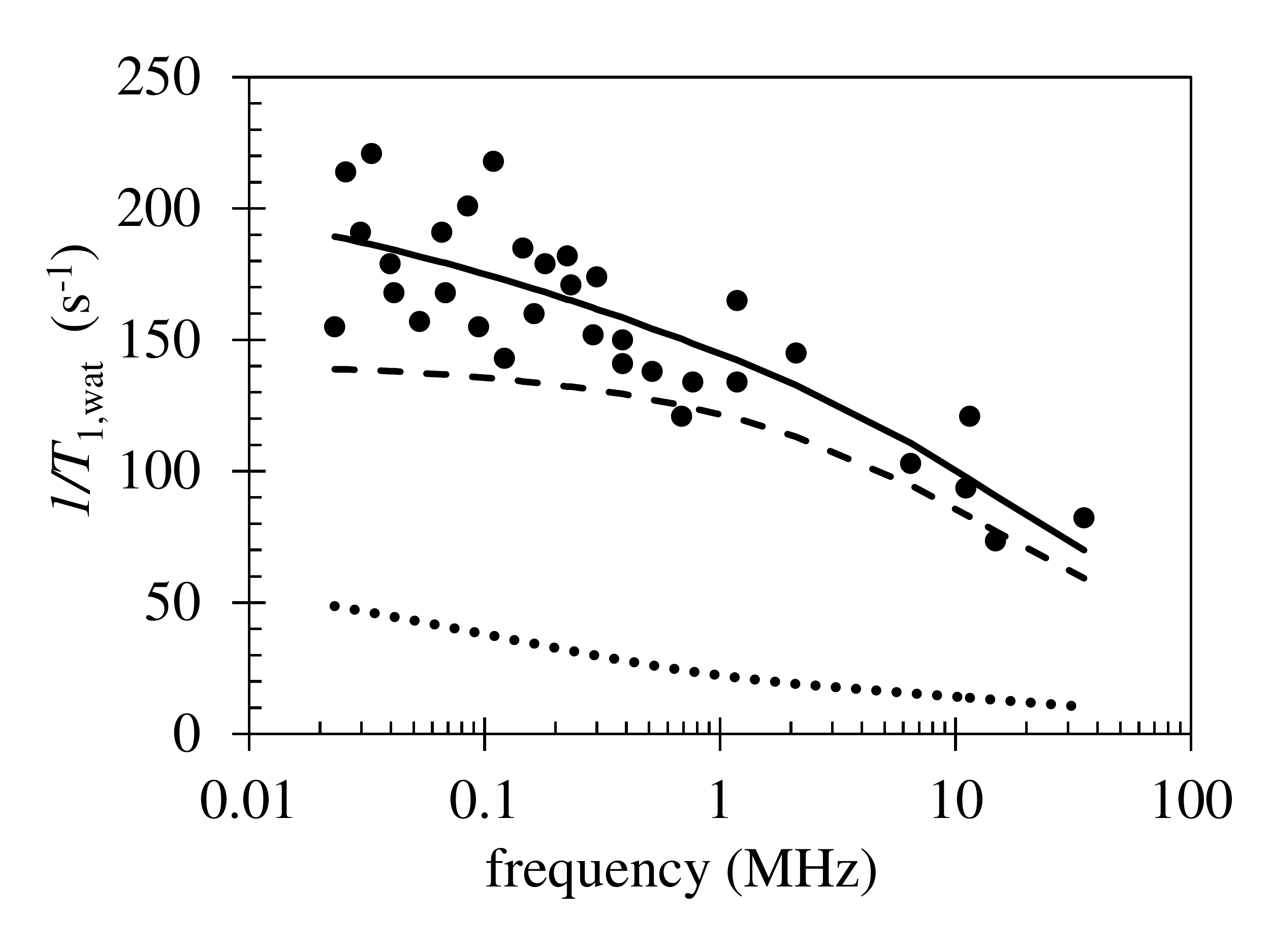} \\[-0.2 cm]
		\caption["Short" caption without tikz code]{$T^{-1}_\onewater$ is presented as a function of frequency for water in an oil shale.   The experimental data ($\bullet$) is from \cite{Korb.2014}. The theoretical curve (\raisebox{.8ex}{\line(1,0){20}}) is composed of a contribution due to the interaction of bulk water with  Mn$^{2+}$ in the pore walls ($\cdots$) and aqueous Mn$^{2+}$ (\raisebox{.8ex}{\line(1,0){5}~~\line(1,0){5}}).\\[-0.8cm]}
		\label{Fig3:shale_water}
	\end{center}
\end{figure}

The model is now fit to the $T_1^{-1}$ dispersions from the first (and only to date) experimental study of an oil shale due to Korb and co-workers \cite{Korb.2014}. The separate oil and water dispersions are presented in Figs.~\ref{Fig2:shale_oil} and \ref{Fig3:shale_water}.  Notice how the two data sets have different functional dependence on frequency indicating different distributions within the pore. Spin relaxation in this oil shale is due to the interaction of $^1$H in the water and oil with Mn$^{2+}$ ions identified as the dominant paramagnetic species by electron spin resonance  \cite{Korb.2014}.  Fits are undertaken by varying $\tau_\bulk$, $\tau_\surf$ and $\tau_\DD$ with fit quality assessed using a simple least-squares measure.
 
\begin{figure}[tbh!]
	\begin{center}
		\includegraphics[width=8.6 cm, trim={1cm -7cm 1cm 13cm},clip]{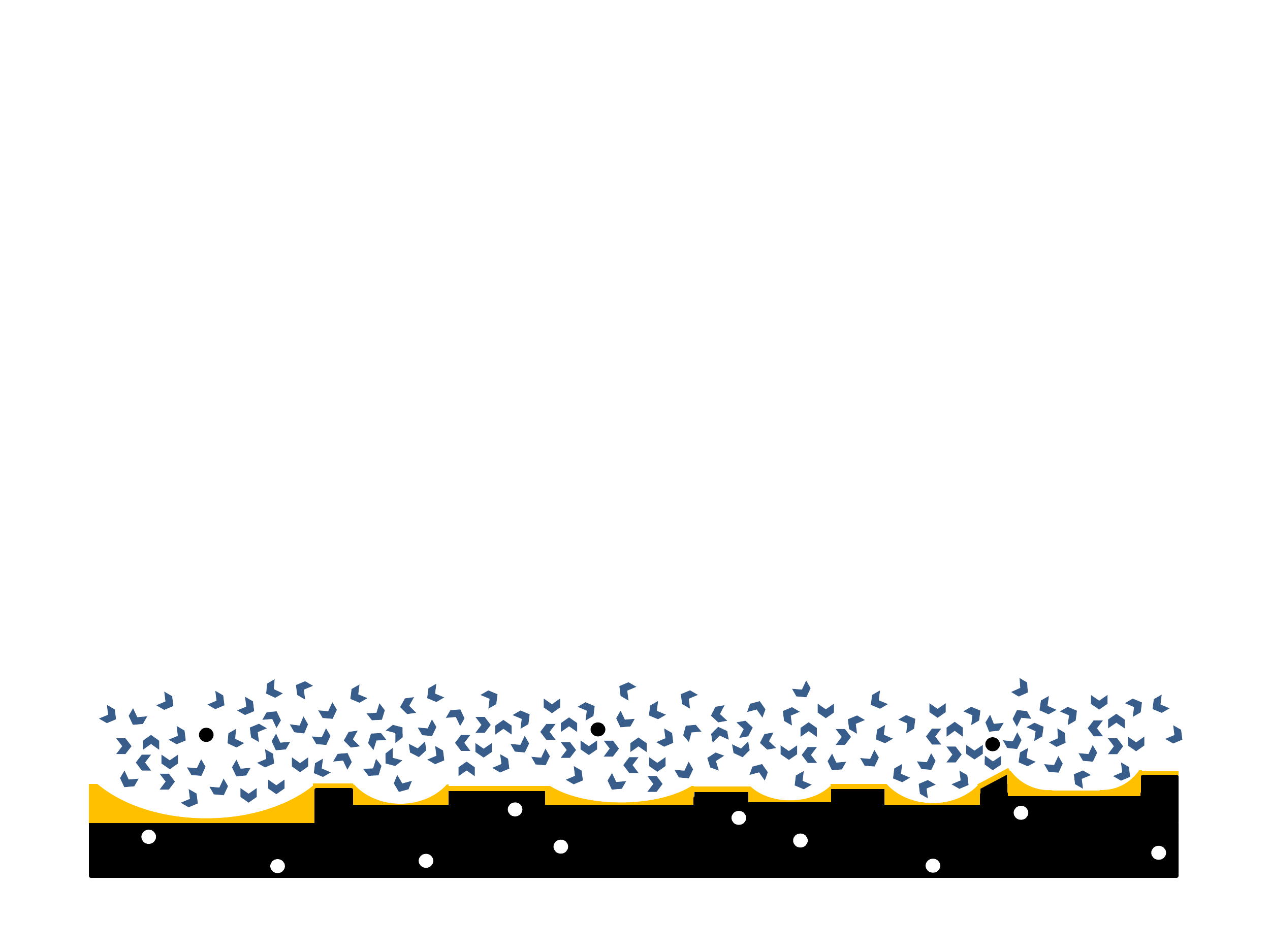} \\[-3.2 cm]
		\caption["Short" caption without tikz code]{A cartoon  of a pore surface consistent with the findings. The pore wall (black) contains rare Mn$^{2+}$ impurities (white circles) with an oil surface (gray).  The water (chevrons) contains  desorbed Mn$^{2+}$ (black circles).\\[-1.0cm]  }
		\label{Fig4_schematic_oil+water}
	\end{center}
\end{figure}

The pore surface is found to be {\em oil} wetting. Contributions to $T_\oneoil^{-1}$ are due to the interaction of Mn$^{2+}$ impurities in the pore walls with surface oil ($T_\onesurfoil^{-1}$) and bulk oil ($T_\onebulkoil^{-1}$).
$G_\bulkoil (t)$ is calculated from Eq.~(\ref{eqn:G4}) using the parameters for bulk oil in Table~\ref{table} and  $G_\surfoil (t)$ may be written
\begin{eqnarray}
G_\surfoil (t) & = &  f\:G(t) + (1\!-\!f)\: G(t) \:e^{-t/\tau_\DD}  \label{eqn:Gshaleoilsurf} \label{Goilsurf}
\end{eqnarray}
where $G(t)$ is calculated using Eq.~(\ref{eqn:G4}) using tabulated parameters for surface oil. Eq.~(\ref{Goilsurf}) could arise if a fraction $f$ of the surface comprises a mono-layer of oil where no desorption occurs over the time scale of $T_1$ or $T_2$.  This would arise with droplets of oil occupying $(1-f)$ of the surface area or in pits (Fig.~\ref{Fig4_schematic_oil+water}).  $T_\oneoil^{-1}$ is then found via
\begin{eqnarray}
\!\!\!\!\!\!\!\!\!\!\!\!T_\oneoil^{-1}(\tau_\surf, \tau_\bulk, \tau_\DD) & = &  x \: T_\onesurfoil^{-1}(\tau_\surf, \tau_\DD) +  (1\!-\!x)\: T_\onebulkoil^{-1} (\tau_\bulk) \label{eqn:T1shaleoil}
\end{eqnarray}
where the explicit dependence on the $\tau$ parameters is indicated.  The quantity $x$ represents the fraction of oil in the surface layer and Eq.~(\ref{eqn:T1shaleoil}) is justified if $\min (T_1,T_2)\! \gg \! \max (\tau_\surf,\tau_\DD,\tau_\bulk)$, the so-called fast-diffusion limit. 

Parameters and fit outcomes are listed in Table~\ref{table}.  Satisfactory fits {\em cannot} be obtained using the experimental Mn$^{2+}$ spin density of $N \! \approx \! 0.5$/nm$^{3}$ \cite{Korb.2014}.  It is found that the effective Mn$^{2+}$  density is about $N/20$. This is not unexpected, a non-linear relationship between relaxation rate and impurity density is well known.  The impact of the Mn$^{2+}$ impurities is reduced due to  clustering in the rock and, as would appear here, desorption of Mn$^{2+}$ at pore walls into the pore fluid.

Results show that $\tau_\DD \! \approx \! \tau_\surf$ suggesting that surface diffusion and desorption of oil molecules are linked processes, very different from the Korb model. A mechanism consistent with this result allows a surface molecule to depart the surface, the vacancy filled by a second surface molecule (rather than by a bulk molecule whose passage is blocked) leaving a second vacancy which is either filled by the desorbed molecule (exchange), a bulk molecule or a another surface molecule.  The mechanism is illustrated in Fig.~\ref{Fig1_model}. It is noted that since surface molecules only execute a few hops before desorbing, the surface only needs to be locally flat for the pore model of Fig.~\ref{Fig1_model} to be valid.

$T_\onesurfoil^{-1}$ was explored for different values of the L\'{e}vy parameter $\alpha$. For $\alpha\!=\!0.5$, $T_\onesurfoil^{-1}$ differs by at most 10\% over the frequency range of fits but overall fit quality is unchanged compared to Fickian statistics with $\alpha\!=\!2$.  This is because the dominant contribution to $T_\onesurfoil^{-1}$ arises for surface spins which make just a few hops on the surface prior to desorption. This contribution is adequately described by Fickian dynamics.  Whilst surface spin diffusion is almost certainly a  L\'{e}vy process, the difference between Fickian and  L\'{e}vy dynamics does not in practice reveal itself in fits to this set of dispersion data. \\[-0.6cm]
 \begin{table}[ht!]
 	\caption{List of model parameters required to fit to the $T_1^{-1}$ dispersion \cite{Korb.2014} for oil and water in oil shale. $\delta\!=\!0.27$~nm.}
 	
 	\begin{tabular}{lll}\toprule
 		Parameter~~~&Oil~~~~~~~~~~~~  & Water~~~~~ \\ \hline
 		 $f$   &   0.1--0.2     &   --    \\
 		 $d_\surf/d_\bulk$   &   2$\delta /3 \delta$     &  --/3$\delta$   \\
 		 $h_\surf/h_\bulk$   &   $\delta/18 \delta$     &  --/18$\delta$     \\
 		  		$\alpha$  &   2 & -- \\
 		$\tau_\bulk$       &       20--40~ps                 &    10--40~ps            \\
 		$\tau_\surf$       &        0.1--0.5~$\mu$s               &     -- \\
 		$\tau_\DD$       &       0.2--0.3$~\mu$s                &     --           \\
 		\hline \hline
 	\end{tabular} 
 	\label{table}
 \end{table}

The bulk oil correlation time $\tau_\bulk$ lies in the range 20-40 ps, consistent with, but slightly  longer than, typical pure alkanes (15 ps) \cite{Blanco.2008}.
The $T_\oneoil/T_\twooil$ ratio, which ranges from 5 to 10 experimentally \cite{Korb.2014}, is found to be a strong function of $\tau_\surf$ with $\tau_\surf\!=\!0.1\mu$s corresponding to $T_\oneoil/T_\twooil\!\approx\!5$ and $\tau_\surf\!=\!0.5\mu$s to $T_\oneoil/T_\twooil\!\approx\!10$.  This result suggests that the $T_\oneoil/T_\twooil$ ratio might provide a direct measure of surface affinity. Combined with peak-spread information, it may be possible to infer oil chain length and surface affinity from $T_1$--$T_2$ maps.  With down-bore $T_1$--$T_2$ mapping a possibility in the future, the significance of this result is obvious.

Analysis of the $T_\onewater^{-1}$ dispersion for water, that is a different shape to oil,  reveals that  $T_\onesurfwater^{-1}$ does not contribute to the measured dispersion and therefore water is {\em not} located on the pore surface -- an independent observation compatible with an oil-wetting shale.  Yet the magnitude of the experimental $T_\onewater^{-1}$ dispersion provides unequivocal evidence of interaction with  Mn$^{2+}$ ions. It is therefore proposed that Mn$^{2+}$ ions are present in the bulk water.    This conclusion is supported by the earlier observation that  Mn$^{2+}$ impurities are depleted at the pore surfaces, presumably having desorbed over millennia into the bulk water.  It is noted that Mn$^{2+}$ was not found in the oil where it is insoluble. It is noted that $T_\onewater/T_\twowater$ for water in oil shale is typically $\approx 2$ \cite{Fleury.2016}, close to that for MnCl$_2$ solution \cite{Pykett.1983}.

The contribution $T_\oneMn^{-1}$ due to aqueous Mn$^{2+}$ is estimated from the expression obtained for bulk water  \cite{Abragam,Faux.1986} adapted to describe the relative motion of water spins with respect to a Mn$^{2+}$ ion assumed to be static.  Therefore
\begin{eqnarray}
T_\onewater^{-1}(\tau_\bulk) & = &  T_\onebulkwater^{-1} (\tau_\bulk) + T_\oneMn^{-1} (\tau_\bulk) \label{eqn:T1shalewater}
\end{eqnarray}
which has a single fit parameter, $\tau_\bulk$.   Optimum fits (Fig.~\ref{Fig3:shale_water}) are obtained for $\tau_\bulk \approx$ 10--40~ps, longer than for pure water at room temperature (5.3~ps) but consistent with a reduction of the diffusion coefficient due to dissolved ions and molecules. The aqueous Mn$^{2+}$ density is found from the fits to be 5--7.5~mM, a factor 100-150 more dilute than the  measured equivalent density in the solid. Assuming pores are mostly water-filled and that all surface Mn$^{2+}$ has desorbed, the mean pore thickness is estimated at 50--80~nm. 

In summary, a general model is proposed which captures the molecular dynamics of fluids in porous solids.  The theory is presented which translates the model to $T_1^{-1}$ dispersions and is tested  by fitting to NMRD measurements on an oil shale.  The analysis yields a wealth of physically-reasonable time constants  which are consistent between the two co-existing fluids, provides insight into diffusion mechanisms and pore morphology. The 3$\tau$ model and theoretical results are applicable to any porous systems containing $^1$H spins in motion relative to fixed paramagnetic impurities and
establishes NMRD as a powerful experimental tool for measuring the dynamical properties of fluids in porous solids.

\bibliography{References}

\end{document}